\documentclass[pra,twocolumn,floatfix,superscriptaddress,longbibliography,notitlepage]{revtex4-1}

\usepackage{soul}
\usepackage{graphicx, color, graphpap}% Include figure files
\usepackage{enumitem}
\usepackage{amssymb}
\usepackage{amsthm}
\usepackage{multirow}
\usepackage[colorlinks=true,citecolor=blue,linkcolor=magenta]{hyperref}
\usepackage[T1]{fontenc}
\usepackage{bbm}
\usepackage{thmtools,thm-restate}
\usepackage{verbatim}
\usepackage{mathtools}
\usepackage{titlesec}
\usepackage{amsmath}
\usepackage[title]{appendix}

\usepackage[caption = false]{subfig}

\long\def\ca#1\cb{} %Use for commenting out: \ca...\cb

\newcommand{\ket}[1]{|#1\rangle}               %ket
\newcommand{\poly}{\operatorname{poly}}

\newcommand{\OC}{\mathcal{O}}

\newcommand{\SC}{\mathcal{S}}

\newcommand{\Tr}{{\rm Tr}}

\newcommand{\Var}{{\rm Var}}

\renewcommand{\geq}{\geqslant}
\renewcommand{\leq}{\leqslant}

\renewcommand{\vec}[1]{\boldsymbol{#1}}  % Bold vectors instead of arrow vectors

\newcommand{\ad}{^\dagger}
\newcommand*{\id}{\openone}

\newcommand{\thv}{\vec{\theta}}

{}%         Body font
{}%         Indent amount (empty = no indent, \parindent 
\newtheorem{theorem}{Theorem}
\newtheorem{lemma}{Lemma}
\newtheorem{corollary}{Corollary}

\newtheorem{definition}{Definition}
\begin{document}
\author{Andrew Arrasmith} 
\affiliation{Theoretical Division, MS B213, Los Alamos National Laboratory, Los Alamos, NM 87545, USA.}
%\affiliation{Quantum Science Center, Oak Ridge, TN 37931, USA}

\author{Zo\"{e} Holmes}
\affiliation{Information Sciences, Los Alamos National Laboratory, Los Alamos, NM USA.}

\author{M. Cerezo}
\affiliation{Theoretical Division, MS B213, Los Alamos National Laboratory, Los Alamos, NM 87545, USA.}
\affiliation{Center for Nonlinear Studies, Los Alamos National Laboratory, Los Alamos, NM, USA
}
%\affiliation{Quantum Science Center, Oak Ridge, TN 37931, USA}

\author{Patrick~J.~Coles} 
\affiliation{Theoretical Division, MS B213, Los Alamos National Laboratory, Los Alamos, NM 87545, USA.}
%\affiliation{Quantum Science Center, Oak Ridge, TN 37931, USA}

\title{Equivalence of quantum barren plateaus to cost concentration and narrow gorges}

\begin{abstract}
Optimizing parameterized quantum circuits (PQCs) is the leading approach to make use of near-term quantum computers. However, very little is known about the cost function landscape for PQCs, which hinders progress towards quantum-aware optimizers. In this work, we investigate the connection between three different landscape features that have been observed for PQCs: (1) exponentially vanishing gradients (called barren plateaus), (2) exponential cost concentration about the mean, and (3) the exponential narrowness of minina (called narrow gorges). We analytically prove that these three phenomena occur together, i.e., when one occurs then so do the other two.  A key implication of this result is that one can numerically diagnose barren plateaus via cost differences rather than via the computationally more expensive gradients. More broadly, our work shows that quantum mechanics rules out certain cost landscapes (which otherwise would be mathematically possible), and hence our results could be interesting from a quantum foundations perspective. 
\end{abstract}

\maketitle
\section{Introduction}

The current availability of cloud-based quantum computers has led to great excitement over the possibility of obtaining quantum advantage for chemistry, materials science, and other applications in the near future. However, these Noisy, Intermediate-Scale Quantum (NISQ) devices are limited by both the number of qubits available and the hardware noise~\cite{preskill2018quantum}. Parameterized quantum circuits (PQCs) have been proposed as an ideal strategy to deal with the practical limitations of NISQ devices. Indeed, PQCs are employed in both variational quantum algorithms~\cite{cerezo2020variationalreview,bharti2021noisy,endo2021hybrid,peruzzo2014variational,farhi2014quantum,mcclean2016theory,khatri2019quantum,sharma2019noise,larose2019variational,arrasmith2019variational,cerezo2020variationalfidelity,endo2020variational,cirstoiu2020variational} and quantum neural networks~\cite{schuld2014quest,cong2019quantum,verdon2018universal,abbas2020power,beer2020training,biamonte2017quantum}, which respectively focus on ground state (and related) applications and data classification applications. PQCs have the potential to solve problems with shorter circuits and fewer qubits than traditional approaches.

When training PQCs, one efficiently evaluates a task-specific cost function on a quantum computer, while employing a classical optimizer to optimize the PQC parameters. Off-the-shelf classical optimizers may not perform as well as optimizers that exploit the nature of quantum cost function landscape, and this has led to the field of quantum-aware optimizers~\cite{kubler2020adaptive,arrasmith2020operator,stokes2020quantum,koczor2019quantum,sweke2020stochastic,nakanishi2020sequential,harrow2019low,lavrijsen2020classical,parrish2019jacobi,fontana2020optimizing}. Of course, understanding the cost function landscape for PQCs will be crucial for designing quantum-aware optimizers. Unfortunately, very little is known about such landscapes. What makes quantum landscapes different from landscapes encountered in classical optimization? This is a fundamental question, not only for practical applications of quantum computers, but also for quantum foundations (i.e., efforts to understand the uniqueness of quantum theory relative to other mathematical theories~\cite{janotta2014generalized}).

One important discovery about these landscapes is that deep PQCs (i.e., those that have many sequential quantum operations) exhibit \textit{barren plateaus}~\cite{mcclean2018barren}. A barren plateau (BP) is a landscape where the magnitudes of gradients are exponentially suppressed with growing problem size. This result was generalized to show that the higher the expressibility of a PQC (i.e., how many different states it can prepare), the more the gradients will be suppressed~\cite{holmes2021connecting}. It was also proven that when the cost function depends on global properties of the solution state,  BPs arise even for shallow PQCs~\cite{cerezo2020cost,uvarov2020barren}. Large degrees of entanglement can also give rise to BPs~\cite{sharma2020trainability,marrero2020entanglement,patti2020entanglement}. Finally, BPs can arise due to quantum error processes washing out all landscape features, and these BPs are called noise-induced BPs~\cite{wang2020noise}. Due to the prevalence of BPs, several strategies have been developed to circumvent or mitigate the effect of barren plateaus~\cite{verdon2019learning,volkoff2021large,skolik2020layerwise,grant2019initialization,pesah2020absence,zhang2020toward,bharti2020quantum,cerezo2020variational,sauvage2021flip,liao2021quantum}.

A different landscape feature that has been observed for PQCs is a \textit{narrow gorge}~\cite{cerezo2020cost}. The narrow gorge phenomenon is where the well around a minimum contracts exponentially quickly with growing system size. The narrow gorge phenomenon is not fully general, as for noise-induced BPs the minima are flattened as well. However, recognizing that  narrow gorges are caused by the cost function values probabilistically concentrating about a mean, we can rephrase this phenomenon in terms of concentration. It has been suggested previously that the exponential concentration of the cost values (often in the form of a narrow gorge) always accompanies a BP, but this has not yet been proven~\cite{cerezo2020cost}.

\begin{figure*}[t!]
    \centering
    \includegraphics[width=0.75\textwidth]{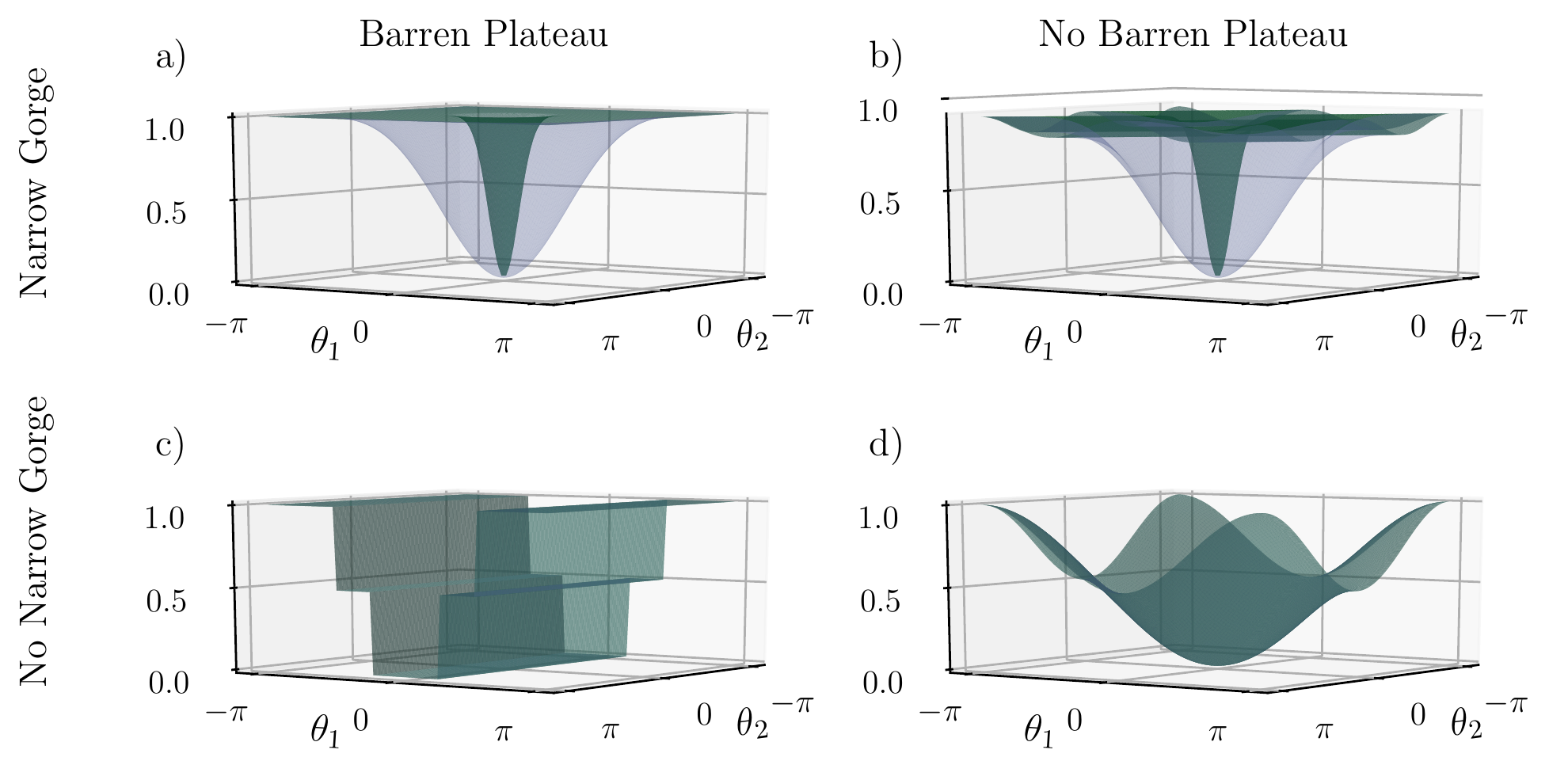}
    \caption{\textbf{Schematic representation of our results.} Here we plot in green (blue) a cross section of four different theoretically possible cost landscapes for a system with $n = 50$ ($n=4$) qubits. Landscape a) has a narrow gorge and a barren plateau, landscape b) has a narrow gorge but no barren plateau, landscape c) has no narrow gorge but a barren plateau, and landscape d) has no narrow gorge and no barren plateau. While these four variations are mathematically possible, we prove that for parameterized quantum circuits (PQCs), under a natural set of assumptions, barren plateaus and narrow gorges are perfectly correlated. That is, for PQCs, only options a) and d) are possible. For details of the cost functions plotted here, see Appendix~\ref{sec:landscapes}.}\label{fig:Schematic}
\end{figure*}

In this work, we study a general class of PQCs and cost functions widely used in the literature and we prove that the suppression of cost function gradients is always accompanied by a probabilistic concentration of cost values, and vice versa. As a consequence, we show that saying that a landscape features a BP is logically equivalent to saying that the cost values are, on average, exponentially concentrated about the mean value, which for landscapes with a well-defined minimum means a narrow gorge. Figure~\ref{fig:Schematic} illustrates this phenomenon. This result is practically significant as it means that numerical testing for a barren plateau can be significantly sped-up by gathering statistics on finite differences between random points rather than computing costly gradients. At the conceptual level, our results shows that quantum mechanics rules out certain kinds of cost landscapes (see Fig.~\ref{fig:Schematic}), which may have foundational interest.

Below we first provide background regarding PQCs and barren plateaus. Next we present our analytical results proving the connection between different landscape features for PQCs and linear cost functions (i.e., cost functions where the one computes the expectation value of Hermitian operators with the respect to the PQCs output state). We then show a numerical demonstration of the similar scaling of the variances of finite differences and gradients. Finally, we conclude and discuss the practical applications of this result.

\section{Background}\label{sec:Background}

To set the stage for our analytical results we will first review some prior results and establish our notation. First, we will define the class of cost functions we consider and state our assumptions about the ansatzes used. Next we discuss barren plateaus and state the formal definition we will be working with. We then review a recent result that ties the expressibility of a quantum circuit to the suppression of cost function gradients. Finally, we discuss previous results on the narrow gorge phenomenon and more generally the concentration of cost function values.

\subsection{Cost Functions and PQC Structure}\label{sec:costs}

In this work we will consider a highly general form of a cost function with which we will train our parameterized quantum circuit to minimize. The PQC is expressed as a unitary $U(\thv)$ parameterized by a vector $\thv$ of $m$ continuous parameters. 
If we are optimizing over a training set $\SC$ containing $S$ initial state and measurement operator pairs $(\rho_i, O_i )$, we will write this cost as
\begin{equation}\label{eq:cost}
    C(\thv) =\sum_{i=1}^{S}a_i \Tr[U(\thv)\rho_i U^\dagger(\thv) O_i].
\end{equation}
Here $a_i$'s are just the coefficients of the linear combination of expectation values. The subscript $i$ highlights the fact that one could work with different functions for each state in the training set. We note that for VQAs
such as the variational quantum eigensolver (VQE) the training set would be of size $S=1$. For quantum machine learning tasks, however, one will typically work with larger training sets.

Without loss of generality, we will write the parameterized unitary $U$ in the following form:
\begin{equation}\label{eq:PQC2}
    U(\thv) = \prod_{l=1}^{m} e^{-i\theta_lH_l} W_l  \, ,
\end{equation}
where the $\{W_l\}$ is some set of fixed unitaries. We further assume that the generators $H_l$ have two distinct, non-zero eigenvalues normalized to be $\pm 1$,  such that $H_l$ is an involutory operator, i.e., $H_l^2=\id$. In this case, it suffices to choose the parameters $\theta_l\in[0,2\pi]$ as $e^{-i\theta_lH_l}=e^{-i(\theta_l+2\pi)H_l}$, and hence the parameter space is periodic and has a maximum length scale $L_\textrm{max}$ which is in $\OC(\poly(m))$. Finally, for the sake of computational efficiency we assume that for $n$ qubits the number of parameters $m$ is in $\OC(\poly(n))$.

We will later wish to consider the portions of the PQC $U(\thv)$ that come before or after a given parameter $\theta_j$. To that end we define the left and right portions of the PQC split at the index $j$ to be 
\begin{equation}
    U(\thv)_{L,j}=\prod_{l=j+1}^{m} e^{-i\theta_lH_l/2} W_l\,,
\end{equation} 
and \begin{equation}
    U(\thv)_{R,j}=\prod_{l=1}^{j} e^{-i\theta_lH_l/2} W_l\, ,
\end{equation} 
respectively.

For the sake of completeness we here review the assumptions we have made. First, we consider linear cost functions such as those in Eq.~\eqref{eq:cost}, where the $C(\thv)$ is expressed as a weighted summation of expectation values. While more general functions are also used in the literature (such as log-likelihood cost functions), we note that the results presented in this manuscript can also, in many cases, be extrapolated beyond the linear functions considered here. We refer the reader to Ref.~\cite{thanasilp2021subtleties} for further discussion. Second, we assume that the PQC in Eq.~\eqref{eq:PQC2} is generated by involutory operators $H_l$ that only have two distinct eigenvalues. This is quite general as most quantum hardware work with basic gates obtained by the exponentiation of Pauli operators, which are involutory. Then we assume that the PQC has a number of parameters that grows at most polynomially with the system size. This is necessary as circuits with more parameters (e.g. $m\in\OC(2^n)$) will be inefficient to train. Finally, we have assumed that the parameters in $U(\thv)$ are chosen independently, which is a standard assumption in the literature.

\subsection{Barren Plateaus}
A barren plateau landscape is one where the average magnitudes of the cost gradients are exponentially suppressed. This notion is usually mathematically formalized in terms of the mean and variance of partial derivatives of the cost function.
\begin{definition}[Barren Plateau]\label{def:BP}
Consider the cost function defined in Eq.~\eqref{eq:cost}. This cost exhibits a barren plateau if, for all $\theta_\mu\in\thv$, the variance of its partial derivative vanishes exponentially with $n$, i.e., as
\begin{equation}\label{eq:var}
    \Var_{\thv}[\partial_\mu C(\thv)]\leq F(n)\,,\quad \text{with}\quad F(n)\in\OC\left(\frac{1}{b^n}\right)\,.
\end{equation}
for some constant $b> 1$. As indicated, the expectation values are taken over the parameters $\thv$.
\end{definition}

An immediate consequence of Definition~\ref{def:BP} is that the fraction of parameter space with a partial derivative magnitude above a threshold $c>0$ is also exponentially suppressed. To see this, first note that on a periodic parameter space the mean partial derivative $\text{E}_{\thv}[\partial_\mu C(\thv)]$ will always be exactly zero as this follows from the fact that the integral of $\partial_\mu C(\thv)$ along a closed curve must be zero.
It then follows from  Chebyshev's inequality, that the probability that the cost function partial derivative deviates from its mean value (of zero) is bounded as
\begin{equation}\label{eq:cheb}
    P(|\partial_\mu C(\thv)|\geq c)\leq \frac{\Var_{\thv}[\partial_\mu C(\thv)]}{c^2}\,.
\end{equation}
Thus if the variance in the partial derivative of the cost vanishes exponentially, the probability that the partial derivative
is non-zero similarly vanishes.

When a cost exhibits a barren plateau according to Definition~\ref{def:BP}, an exponentially large precision (i.e., an exponentially large number of shots) is needed to navigate through the flat landscape and determine a cost minimizing direction. Moreover, the exponential precision requirement has been shown to affect both derivative-based~\cite{cerezo2020impact} and derivative-free~\cite{arrasmith2020effect} optimization methods, meaning that one cannot improve the trainability of the cost function in a barren plateau by simply changing the optimization method.

The barren plateau phenomenon was first identified in~\cite{mcclean2018barren}. Therein, the authors showed that randomly initialized deep unstructured circuits exhibit exponentially vanishing gradients. Here, the mechanism leading to barren plateaus is essentially that the set of unitaries generated by the parameterized quantum circuit is random enough that it becomes a $2$-design (i.e. the distribution matches up to the second moment that of the uniform Haar distribution of unitaries)~\cite{brandao2016local,dankert2009exact}. In this case, the randomness of the circuit scrambles the information processed by the PQC and makes the cost function values concentrate around their average, and hence changes in individual parameters lead to small changes in cost values. This intuition has also been used to show that barren plateaus preclude learning scramblers~\cite{holmes2020barren}. In addition, we note that the randomness-induced barren plateaus can also be linked to the entanglement generated by the quantum circuit. Specifically, it has been shown that circuits which generate large amounts of entanglement also lead to barren plateaus~\cite{sharma2020trainability,patti2020entanglement,marrero2020entanglement}.

The barren plateau phenomenon was later studied in shallow layered hardware efficient PQC which are not random enough (or entangling enough) to lead to randomness-induced barren plateaus. In~\cite{cerezo2020cost} the authors analyze parametrized quantum circuits that are composed of random two-qubit gates acting on neighboring qubits in a brick-like fashion. Here it was proved that the locality of the operators $O_i$ in Eq.~\eqref{eq:cost} can be linked to the presence of barren plateaus, so that global cost functions (i.e. when $O_i$ acts non-trivially on all qubits) have barren plateaus irrespective of the circuit depth. In this case, the mechanism leading to barren plateaus is not the randomness in the circuit, but rather the fact that the cost function trains the parameters by comparing objects that live in exponentially large Hilbert spaces. On the other hand, it was shown that local cost functions (i.e. when $O_i$ acts non-trivially on at most two neighboring)  can be generically trainable for shallow circuits as here the variance  $\Var_{\thv}[\partial_\mu C(\thv)]$ is at worst polynomially vanishing with $n$.  The connection between the locality of the cost function and the presence of barren plateaus has also been extended to the context of  tensor-network based machine learning models in~\cite{liu2021presence}. Therein, it was shown that global cost functions also lead to barren plateaus, while local cost functions do not exhibit them.

Finally, we remark that it has also been shown that some types of hardware noise acting throughout the quantum circuit can also lead to barren plateaus. As proven in~\cite{wang2020noise} the effect of unital quantum noise corrupts the quantum state as it evolves through the circuit and maps it to the fixed point of the noise model (i.e., the maximally mixed state). This noise-induced barren plateau then flattens the whole landscape and again leads to exponentially vanishing gradients.

\subsection{Expressibility}

Recent work has generalized the notion of a barren plateau by providing an upper bound on the variance of partial derivatives based on more general notions of expressibility~\cite{holmes2021connecting}. To formalize the notion of expressibility, we define $\mathbb{U}$ as the set of unitaries that are reachable by varying the parameters of the PQC $U(\thv)$. Intuitively, a maximally expressive PQC could reach any unitary in the unitary group $\mathcal{U}(2^N)$. The expressibility is therefore defined by comparing the difference between a uniform distribution over $\mathbb{U}$ and the uniform (Haar) distribution over $\mathcal{U}(2^N)$. This comparison is done by defining the following expressibility super-operator~\cite{sim2005best,nakaji2020expressibility}
\begin{equation}
    \begin{aligned}
        \mathcal{A}_{\mathbb{U}}^{(t)}(\cdot)=&\int_{\mathcal{U}(2^N)} d\mu(V) V^{\otimes t}(\cdot)(V^{\dagger}){\otimes t}\\
        &-\int_{\mathbb{U}} dU U^{\otimes t}(\cdot)(U^{\dagger}){\otimes t}.
    \end{aligned}
\end{equation}
Here $d\mu(V)$ is the Haar measure and $dU$ is the uniform measure over $\mathbb{U}$. If  $\mathcal{A}_{\mathbb{U}}^{(t)}(O)=0$ for all operators $O$, then the $t$-th moments of $\mathbb{U}$ match the value for the Haar distribution and $\mathbb{U}$ is called a $t$-design~\cite{divincenzo2002quantum,gross2007evenly,roberts2017chaos,low2010pseudo,hunter2019unitary}. As a $2$-design is sufficient to guarantee a barren plateau, bounding the suppression of cost gradients only requires considering $t=2$ and we will focus on that case. 

In the context of optimizing a cost of the form in Eq.~\eqref{eq:cost}, we do not need to consider the expressibility on all possible operators but rather on the initial states $\{\rho_i\}$ and on the measurement operators $\{O_i\}$. As such, the quantities
\begin{equation}
    \varepsilon_{\mathbb{U}}^{O_i}\equiv\|\mathcal{A}_{\mathbb{U}}^{(2)}(O_i^{\otimes 2})\|_2
\end{equation}
and
\begin{equation}
    \varepsilon_{\mathbb{U}}^{\rho_i}\equiv\|\mathcal{A}_{\mathbb{U}}^{(2)}(\rho_i^{\otimes 2})\|_2
\end{equation}
capture the expressibility of the PQC relative to the cost function.  
When these quantities are small, the PQC is highly expressible.

The variance of the cost function partial derivative is then bounded from above by~\cite{holmes2021connecting}:
\begin{equation}\label{eq:express_bound}
\begin{aligned}
    \Var_{\thv}[\partial_\mu C(\thv)]\leq& \widetilde{F}(n)+\sum_i\left( 4\varepsilon^{O_i}_{\mathbb{U_L}} \varepsilon^{\rho_i}_{\mathbb{U_R}}\right.\\
    &\left.+ \frac{2^{n+2}\left(\varepsilon^{O_i}_{\mathbb{U_L}}\|O_i\|^2_2+ \varepsilon^{\rho_i}_{\mathbb{U_R}}\|\rho_i\|^2_2\right)}{2^{2n}-1}\right) \, .
\end{aligned}
\end{equation}
Here $\mathbb{U_L}$ and $\mathbb{U_R}$ are the set of unitaries reachable by $U_L(\theta)$ and $U_R(\theta)$, respectively, %We note that for a fully expressive PQC (i.e. a $2$-design), $\widetilde{F}(n)\to F(n)$ and this bound becomes the same as the one in Equation~\eqref{eq:var}.
and $\widetilde{F}(n)\in\OC\left(1/\widetilde{b}^n\right)$ is the variance of the partial derivative of the cost for a 2-design (i.e., a maximally expressive PQC)~\cite{mcclean2018barren}. 
We note that for a maximally expressive PQC (i.e., a $2$-design) this bound becomes an equality. More generally, the bound implies that highly expressive PQCs will have flatter landscapes and hence be harder to train.

\subsection{Narrow Gorges and Cost Concentration}

The notion of a narrow gorge arose alongside investigations of barren plateaus~\cite{cerezo2020cost}. For an example barren plateau landscape considered in that work, the authors found that the volume of the "gorges" near the minima were exponentially small. More precisely, the probability that the cost associated with a randomly sampled parameter vector $\thv$ was lower than any fixed (non-maximal) cost threshold could be upper bounded by a function that was exponentially suppressed with growing numbers of qubits $n$. Motivated by that example, we give the following intuitive definition for a narrow gorge.
\begin{definition}[Narrow Gorge]\label{def:NG}
Consider the cost function defined in Eq.~\eqref{eq:cost}. This cost exhibits a narrow gorge if:
\begin{itemize}
    \item There exist cost minima which are lower than the mean cost value by at least
    \begin{equation}
        \Delta(n)>0\, \textrm{with} \, \Delta(n)\in\Omega\left(1/\textrm{poly}(n)\right).
    \end{equation}
    \item The probability that the cost function value at a point chosen from the uniform distribution over the parameter space differs from the mean by at least $\delta$ is bounded as
    \begin{equation}\label{eq:ng_prob}
    P(|E_{\thv}[C(\thv)]-C(\thv)|\ge \delta)\le \frac{G(n)}{\delta^2}
    \end{equation}
    with $G(n)\in\OC\left(\frac{1}{b^n}\right)$ for some $b> 1$.
\end{itemize}
\end{definition}
We remark that the probability in Eq.~\eqref{eq:ng_prob} is with respect to the uniform distribution over the parameters $\vec{\theta}$. We can therefore understand it as a fractional volume of parameter space. A landscape with a narrow gorge is then one where the cost function departs from the mean value by at least $\delta$ for an exponentially small fraction of the volume of the landscape.

The caveat that there exist cost minima lower than the mean here does not always hold. In addition to simple cases like constant cost functions, we note that this can arise in the case of a noise induced barren plateau, where the noise can wash out the minima~\cite{wang2020noise}. In these cases there may be no gorge, but one can instead consider the concentration of the cost function about the mean. A more general concept than a narrow gorge is then  an  exponential (probabilistic) concentration about the mean. We will define this concentration by bounding the variance of cost function differences as
\begin{equation}\label{eq:concentrate}
    \Var_{\vec{\theta}_A}\left[E[C(\vec{\theta})]-C(\vec{\theta}_A) \right]\le G(n)
\end{equation}
again with $G(n)\in\OC\left(\frac{1}{b^n}\right)$ for some $b> 1$. We note that so long as the minima are at least $\Delta(n)>0$ below the mean cost value for  $\Delta(n)\in\Omega\left(1/\textrm{poly}(n)\right)$, such a concentration implies a narrow gorge via Chebyshev's inequality.

\section{Results}
We begin our results by stating the following Lemma.
\begin{lemma}\label{lem:zero_mean_diffs}
For any cost function defined on a periodic parameter space, the following statements hold.
\begin{itemize}
    \item The mean value of the difference in cost values between $\vec{\theta}_A$, a random draw from the uniform probability distribution over the parameter space, and the point $\vec{\theta}_A+L\hat{e}$ for a deterministically chosen distance $L$ and direction indicated by the unit vector $\hat{e}$ is zero:  
    \begin{equation}
        \text{E}_{\vec{\theta}_A}\left[C(\vec{\theta}_A+L\hat{e}))-C(\vec{\theta}_A)\right]=0.
    \end{equation}
    \item The mean value of the difference in cost values between two points $\vec{\theta}_A$ and $\vec{\theta}_B$, both random draws from the uniform distribution over the parameter space, is zero:  
    \begin{equation}
    \begin{aligned}
        \text{E}_{\vec{\theta}_A}\left[C(\vec{\theta}_B)-C(\vec{\theta}_A)\right]=&\text{E}_{\vec{\theta}_B}\left[C(\vec{\theta}_B)-C(\vec{\theta}_A)\right]\\
        =&0.
    \end{aligned}
    \end{equation}
\end{itemize}

\end{lemma}
We note that Lemma~\ref{lem:zero_mean_diffs} can be thought of as a direct consequence of the fact that the average gradient vanishes, since a partial difference can be computed via $C(\vec{\theta}_B)-C(\vec{\theta}_A)=\int_{\vec{\theta}_A}^{\vec{\theta}_B}\vec{\nabla}C(\thv)\cdot d\thv$. For more details see the proof in Appendix~\ref{app:zero_mean_diffs}. 

\subsection{Barren plateaus imply cost concentration}

We now state the first of our main results.

\begin{theorem}\label{thm:grad->diffs}
For any parameterized quantum circuit as defined in Section~\ref{sec:Background}, if $\Var{}\Big(\partial_{\theta_j}C(\theta)\Big)\le F(n)$ then:
\begin{itemize}
    \item The variance of cost function differences between a randomly chosen point in parameter space, $\vec{\theta}_A$, and the point $\vec{\theta}_A+L\hat{e}$ for any deterministically chosen distance $L$ and direction indicated by the unit vector $\hat{e}$ is then bounded above by:
    \begin{equation}
        \Var_{\vec{\theta}_A}\left(C(\vec{\theta}_A+L\hat{e})-C(\vec{\theta}_A) \right)\le m^2 L^2 F(n)
    \end{equation}
    where $m$ is the dimension of the parameter space.
    \item The variance of the cost function differences between two independently chosen random points in parameter space, $\vec{\theta}_A$ and $\vec{\theta}_B$, is bounded above by:
    \begin{equation}
        \Var_{\vec{\theta}_A}\left(C(\vec{\theta}_B)-C(\vec{\theta}_A) \right)\le m^2 L_{\textrm{max}}^2 F(n).
    \end{equation}
     where $L \leq L_\textrm{max}$ is an upper bound on $L$. 
\end{itemize}
\end{theorem}
As with Lemma~\ref{lem:zero_mean_diffs} the basic idea behind this theorem is that the second moment of a finite difference can be related to the second moment of a gradient by integration. See Appendix~\ref{app:BP->NG} for the full proof of Theorem~\ref{thm:grad->diffs}. As an immediate consequence of this theorem we then have the following corollary.
\begin{corollary}\label{corr:BP=>NG}
If a cost function landscape exhibits a barren plateau, then it also exhibits an exponential concentration of cost values. Additionally, if there exist minima at least $\Delta(n)>0$ below the mean cost value for  $\Delta(n)\in\Omega\left(1/\textrm{poly}(n)\right)$, that landscape also has a narrow gorge.
\end{corollary}
Corollary~\ref{corr:BP=>NG} follows from plugging the Definition~\ref{def:BP} into Theorem~\ref{thm:grad->diffs}, averaging over $\thv_B$, and applying Chebyshev’s inequality. 

We note that combining Theorem~\ref{thm:grad->diffs} and Eq.~\eqref{eq:express_bound} allows one to show the extent to which the cost must be concentrated for a given level of expressibility. Hence, by leveraging the results of Ref.~\cite{holmes2021connecting}, we can connect the expressibility of a PQC to the concentration of cost landscape and also to the existence of narrow gorges. 

We now remark on the significance of this result for the different types of barren plateau landscapes. For barren plateaus arising due to expressibility~\cite{mcclean2018barren,holmes2021connecting}, entanglement~\cite{sharma2020trainability,marrero2020entanglement,patti2020entanglement}, or a global cost function~\cite{cerezo2020cost}, the flatness of the landscape does not preclude good minima. Therefore, assuming a reasonable PQC structure, each of these barren plateau classes will always be accompanied by narrow gorges. Noise-induced barren plateau landscapes~\cite{wang2020noise}, on the other hand, flatten all features (including the minima) exponentially quickly and so naturally exhibit cost concentration. However, barren plateau landscapes due to noise do not exhibit narrow gorges as they do not have any gorges left to be narrow.

\subsection{Cost concentration implies barren plateaus}

We also have the following result.
\begin{lemma}\label{lemma:diffs->grad}
For any parameterized quantum circuit as defined in Section~\ref{sec:Background}, if $\Var_{\vec{\theta}_A}\left(C(\vec{\theta}_A+L\hat{e})-C(\vec{\theta}_A) \right)\le G(n)$ for any deterministically chosen distance $L$ and direction indicated by the unit vector $\hat{e}$ then the variance of the gradient can be bounded above by:
\begin{equation}
    \Var{}\Big(\partial_{\theta_j}C(\theta)\Big)\le \frac{G(n)}{4}.
\end{equation}
\end{lemma}
This lemma follows directly from noting that the Parameter Shift Rule~\cite{mitarai2018quantum,schuld2019evaluating}  allows us to write
\begin{equation}\label{eq:PSR}
    \partial_{\theta_j}C(\thv)=\frac{C(\thv'+\pi\hat{e}_j)-C(\thv')}{2}
\end{equation}
where $\hat{e}_j$ is the unit vector along the $j-th$ parameter direction and $\thv'=\thv-\pi/2\hat{e}_j$. We refer the reader to Appendix~\ref{app:E} for a derivation of Eq.~\eqref{eq:PSR}.

Combining Theorem~\ref{thm:grad->diffs} and Lemma~\ref{lemma:diffs->grad} then gives us the following result. 
\begin{theorem}\label{thm:grad=diffs}
For any parameterized quantum circuit as defined in Section~\ref{sec:Background} the following statements hold
\begin{itemize}
    \item A cost landscape is a barren plateau landscape if and only if it exhibits exponential concentration of the cost function as in Eq.~\eqref{eq:concentrate}.
    \item A cost landscape for which there exist minima at least $\Delta(n)>0$ below the mean cost value (with  $\Delta(n)\in\Omega\left(1/\textrm{poly}(n)\right)$) is a barren plateau landscape if and only if it is a narrow gorge landscape. 
\end{itemize}
\end{theorem}
As we will demonstrate numerically in the following section, Theorem~\ref{thm:grad=diffs} implies that when numerically testing for barren plateaus, one can look at the variance of cost differences rather than computing the variance for comparably costly gradient evaluations. This result allows for a significant speed-up for numerically testing for barren plateaus since it removes the need to consider each parameter separately. 

\section{Numerical Demonstration}\label{sec:Numerics}
Here we numerically demonstrate the equivalence of cost concentration and barren plateaus. For these numerical implementations, we consider a layered hardware efficient PQC,
\begin{equation} \label{eq:HEA}
    U(\thv, D) := \prod_{l=1}^D W V(\thv_l) \, ,
\end{equation}
composed of $D$ alternating layers of random single qubit gates and entangling gates. 
The single-qubit layer consists of a product of random $x$, $y$ and $z$ rotations on each qubit. That is,
\begin{equation}
    V(\thv_l) = \prod_{j=1}^n R_{j}^x(\theta_l^{xj}) R_{j}^y(\theta_l^{yj}) R_{j}^z(\theta_l^{zj})   \, ,
\end{equation}
where $R_j^{k}(\theta_l^{kj})$ is a rotation of the $j_{\rm th}$ qubit by an angle $\theta_l^{kj}$ about the $k = x, y$ or $z$ axis.
The entangling layer,
\begin{equation}
    W = \prod_{j = 1}^{n-1} \text{C-Phase}_{j, j+1} \, ,
\end{equation}
consists of a ladder of controlled-phase operations, $\text{C-Phase}$, between adjacent qubits in a 1-dimensional array. For concreteness, we suppose the circuit is initialized in $\rho = |\psi_0\rangle \langle \psi_0 |^{\otimes n}$ where $\ket{\psi_0} = \exp(-i(\pi/8)\sigma_Y) \ket{0}$ and focus on a 2-local cost where the measurement operator is composed of Pauli-$z$ measurements on the first and second qubits, $O = \sigma^z_1 \sigma^z_2$.

\begin{figure}[t]
    \centering
    \includegraphics[width=\columnwidth]{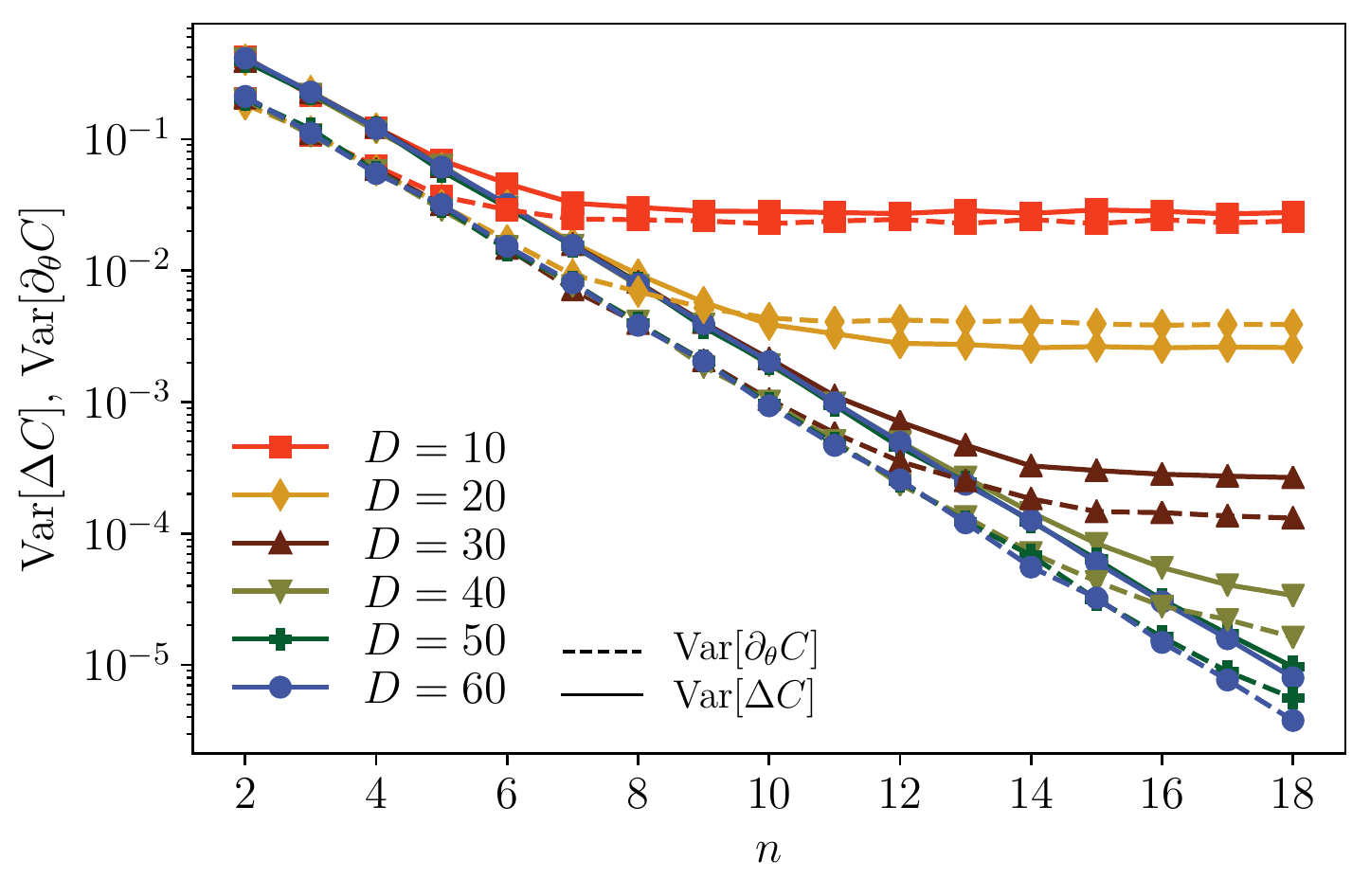}
    \caption{\textbf{Comparing the second moments of finite differences and derivatives.} The variance in the partial derivative of the cost (dashed) and the variance in its finite differences (solid) as a function of the number of qubits $n$ as we vary the circuit depth $D$ of a hardware efficient PQC. In both cases we consider a local cost with $H = \sigma^z_1 \sigma^z_2$, we suppose the computer is prepared in the state $\exp(-i(\pi/8)\sigma_Y) \ket{0}$, and the variance is taken over an ensemble of 2000 random choices in the parameters $\vec{\theta}$ of the layered hardware efficient PQC, Eq.~\eqref{eq:HEA}. The partial derivative is taken with respect to $\theta_1^{x1}$, the rotation angle of the first qubit in the first layer.}
    \label{fig:numerics}
\end{figure}

We used TensorFlow Quantum~\cite{broughton2020tensorflow} to calculate the second moment of the derivative of the cost and its finite differences for an ensemble of 2000 random initializations. In Fig.~\ref{fig:numerics} we plot the variance in the partial derivative with respect to $\theta_1^{x1}$, the $x$ rotation angle for the first qubit in the first layer. While similar behaviour is expected for other parameters (see Appendix~\ref{app:paramdep} for more details) the decision to show the data for $\theta_1^{x1}$ is somewhat arbitrary. Strictly one would need to evaluate the variance of the partial derivative with respect to all $3 D n$ parameters in order to determine whether a general cost has a barren plateau.

In contrast, evaluating the second moment of the cost requires simply generating two random ensembles of cost evaluations, taking their difference, and evaluating the variance of the resulting ensemble of differences. This is substantially less resource intensive than evaluating the variance of the partial derivatives with respect to all parameters. 

As is shown in Fig.~\ref{fig:numerics}, we find that that the second moment of the derivative of the cost and its finite differences scale equivalently, as is expected from Theorem~\ref{thm:grad=diffs}. Specifically, for sufficiently deep circuits, namely $D \geq 60$, both the variance of the partial derivative and cost differences vanish exponentially with the size of the system $n$; however, for shallower circuits a constant scaling may be observed in the asymptotic regime. This constant scaling arises from the fact that in such regimes the ansatz is far from a 2-design and thus highly inexpressive and highly non-entangling. For a more detailed discussion of this see Refs.~\cite{mcclean2018barren, holmes2021connecting, patti2020entanglement,marrero2020entanglement}.

\section{Discussion}

By minimizing both the number of qubits required and the depth of the circuits that need to be executed, variational quantum algorithms and quantum neural networks may yield a quantum advantage before fault-tolerant quantum devices are available. In these algorithms, a problem-specific cost function is minimized using a classical optimizer to vary the parameters in a parameterized quantum circuit (PQC). While this framework minimizes the quantum resources required to perform a computation, it potentially involves a difficult classical optimization.

One particular impediment to efficient optimization is a barren plateau landscape, where gradients are exponentially suppressed as the problem size scales. These landscapes occur for highly expressive PQCs, for cost functions that measure global properties of quantum states, for PQCs that generate high degrees of entanglement, and for hardware with significant noise. Another problematic landscape feature is a narrow gorge, where the fraction of parameter space below some value becomes exponentially suppressed. While these narrow gorges were originally discovered in connection to barren plateaus, there has been debate as to whether or not they are always connected.

In this work, we introduced a bound for how concentrated a cost function is about the mean based on a bound on the variance of the partial derivatives of the landscape. Additionally, we have pointed out that, due to the parameter shift rule for computing derivatives, it is simple to bound the variance of the partial derivatives by a bound on the variance of finite differences computed at random points. As a consequence of these bounds, we have proven that, assuming minima below the mean values exist, saying that a landscape exhibits a barren plateau is logically equivalent to saying that it exhibits a narrow gorge.

The practical application of our result arises in the process of numerically demonstrating a barren plateau. Previous approaches have either needed to focus on the partial derivative of the cost with respect to some particular parameter(s) (potentially missing different behaviors for different parameters) or chosen to accept the computational cost of gathering statistics on the scaling of full gradient vectors. By proving that the exponential suppression of the cost differences between randomly chosen points guarantees a barren plateau, our result allows one to explore the scaling of the entire gradient for a cost equivalent to that of studying a single partial derivative. Our result thus allows for significantly more efficient testing for barren plateaus.

On a deeper level, one can imagine all of the possible landscapes that are allowed by PQCs, and hence by quantum mechanics. We suggest that this field of research could be called \textit{quantum landscape theory}. Our work shows that some landscapes that are mathematically possible are actually ruled out by quantum mechanics (e.g., landscapes with a narrow gorge but no barren plateau, as illustrated in Fig.~\ref{fig:Schematic}). It is worth considering the analogy to efforts to distinguish quantum theory from general mathematical theories~\cite{janotta2014generalized}, which is similar to how we distinguish quantum landscapes from general mathematical landscapes. Hence, we believe that our work may prove interesting from a foundations perspective.

\section*{Acknowledgments}

AA was supported by the U.S. Department of Energy (DOE), Office of Science, Office of High Energy Physics QuantISED program under Contract No.~DE-AC52-06NA25396. ZH and PJC were supported by the Los Alamos National Laboratory (LANL) ASC Beyond Moore's Law project. MC was  supported by the Center for Nonlinear Studies at LANL. This work was supported by the U.S. DOE, Office of Science, Office of Advanced Scientific Computing Research, under the Accelerated Research in Quantum Computing (ARQC) program.

\bibliography{quantum}
\onecolumngrid

\appendix

\section{Visualization of barren plateaus and narrow gorges}\label{sec:landscapes}

To help visualize barren plateaus and narrow gorges, in Fig.~\ref{fig:Schematic} we sketch four possible landscapes: one with a barren plateau and a narrow gorge, one with a barren plateau but no narrow gorge, one with no barren plateau but a narrow gorge and one with neither a barren plateau nor narrow gorge. 

In Fig.~\ref{fig:Schematic}(a) we plot the landscape for the global cost
\begin{equation}
    C_G := 1  - \prod_{j=1}^n \cos(\theta_j/2)^2 \,,
\end{equation}
which was discussed in Ref~\cite{cerezo2020cost}. This landscape has a barren plateau since the variance of the partial derivative of the cost landscape vanishes exponentially as $\text{Var}\frac{\partial C}{\partial \theta_k} = \frac{1}{8} (\frac{3}{8})^{n-1}$. Similarly, as is visually clear from Fig.~\ref{fig:Schematic}a), the probability that the cost deviates substantially from the mean vanishes exponentially as $2^{-n}$. 

In Fig.~\ref{fig:Schematic}(b) we plot the landscape for a cost of the form 
\begin{equation}
    C_{\rm NG-noBP} = 0.9 C_G - 0.1 C_{\rm osc} \, ,
\end{equation}
where $C_{\rm osc}$ is an oscillatory term which does not scale with the size of the system
\begin{equation}
    C_{\rm osc} = \sin\left(\frac{1}{n}\sum_{j=1}^n \theta_j\right)^2 \, .
\end{equation}
The cost $ C_{\rm NG-noBP}$ inherits the narrow gorge of the global cost $C_G$. However, the presence of the oscillatory term $C_{\rm osc}$ prevents the gradient of the cost (and hence the variance of the gradient of the cost) from vanishing as the size of the system is increased, and therefore $ C_{\rm NG-noBP}$ does not exhibit a barren plateau. 

In Fig.~\ref{fig:Schematic}(c) we plot a cost of the form
\begin{alignat}{2}
C_{\rm BP-noNG} =
\left\{
\begin{array}{ll}
0   & \ \ \  \ \ \  |\frac{2}{n}\sum_{j=1}^{n/2} \theta_j| \leq \pi/4  \\
0.5 & \ \ \  \ \ \   \pi/4 < |\frac{2}{n}\sum_{j=1}^{n/2} \theta_j| < \pi/2 \,. \\
1 & \ \ \  \ \ \ |\frac{2}{n}\sum_{j=1}^{n/2} \theta_j| \geq \pi/2  \ \ \ \ 
\end{array} 
\right.
\end{alignat}
The partial derivative of this cost vanishes for all $\theta_j$ (except for a set of measure zero) for all $n$, and as such the cost trivially has a barren plateau. Concurrently, the probability that the cost deviates from its mean of $5/8$ by more than $1/2$ is equal to $1/4$ for all $n$. Thus the landscape does not have a narrow gorge.

In Fig.~\ref{fig:Schematic}(d) we plot the landscape for the local cost
\begin{equation}
    C_L := 1  - \frac{1}{n}\sum_{j=1}^n \cos(\theta_j/2)^2 \, 
\end{equation}
studied in Ref.~\cite{cerezo2020cost}. The variance of the partial derivative for this landscape vanishes polynomially, with $\text{Var}\frac{\partial C}{\partial \theta_k} = \frac{1}{8 n^2}$, and therefore does not have a barren plateau. Concurrently, the probability that the cost deviates substantially from the mean is approximately constant for large $n$ and therefore this landscape does not have a narrow gorge.
% The independence of this cost landscape on system size is apparent in Fig.~\ref{fig:Schematic}d) where the cross sections of the landscapes perfectly overlap for $n=4$ and $n=50$. 

\section{Proof of Lemma~\ref{lem:zero_mean_diffs}}\label{app:zero_mean_diffs}

We now prove Lemma~\ref{lem:zero_mean_diffs} from the main text. We restate the Lemma here for convenience.
\begin{lemma}
For any cost function defined on a periodic parameter space, the following statements hold.
\begin{itemize}
    \item The mean value of the difference in cost values between $\vec{\theta}_A$, a random draw from the uniform probability distribution over the parameter space, and the point $\vec{\theta}_A+L\hat{e}$ for a deterministically chosen distance $L$ and direction indicated by the unit vector $\hat{e}$ is zero:  
    \begin{equation}
        \text{E}_{\vec{\theta}_A}\left[C(\vec{\theta}_A+L\hat{e}))-C(\vec{\theta}_A)\right]=0.
    \end{equation}
    \item The mean value of the difference in cost values between two points $\vec{\theta}_A$ and $\vec{\theta}_B$, both random draws from the uniform distribution over the parameter space, is zero:  
    \begin{equation}
    \begin{aligned}
        \text{E}_{\vec{\theta}_A}\left[C(\vec{\theta}_B)-C(\vec{\theta}_A)\right]=&\text{E}_{\vec{\theta}_B}\left[C(\vec{\theta}_B)-C(\vec{\theta}_A)\right]\\
        =&0.
    \end{aligned}
    \end{equation}
\end{itemize}

\end{lemma}

We begin our proof by noting that finite differences in the cost function can be written in terms of the integrals:
\begin{equation}
\begin{aligned} \Delta C=&C(\vec{\theta}_B)-C(\vec{\theta}_A)\\
=&\int_{\vec{\theta}_A}^{\vec{\theta}_B}\vec{\nabla}C(\vec{\theta})\cdot d\vec{\theta}\,.
\end{aligned}
\end{equation}

We can integrate along the line segment between $\vec{\theta}_B$ and $\vec{\theta}_A$, setting $\vec{\theta}=\vec{\theta}_A+\ell\hat{\vec{\ell}}$, with $\hat{\vec{\ell}}$ being the unit vector along $\vec{\theta}_B-\vec{\theta}_A$ and $\ell \in [0,L]$ where $L=\|\vec{\theta}_B-\vec{\theta}_A\|$. We then have $d\vec{\theta}=\hat{\vec{\ell}}d\ell$.    To show that the mean difference is zero, we simply take the expectation value of this integral with respect to $\thv_A$
\begin{equation}
\begin{aligned} 
\text{E}_{\vec{\theta}_A}[\Delta C]=&\int_{0}^{L}\text{E}_{\vec{\theta}_A}[\vec{\nabla}C(\vec{\theta}_A+\ell\hat{\vec{\ell}})\cdot\hat{\vec{\ell}}]d\ell \\
=&\int_{0}^{L}\text{E}_{\vec{\theta}_A'}[\vec{\nabla}C(\vec{\theta}_A')\cdot\hat{\vec{\ell}}]d\ell \\
=&0\,.
\end{aligned}
\end{equation}
Here $\vec{\theta}_A'=\vec{\theta}_A+\ell\hat{\vec{\ell}}$, and we have used the fact that
\begin{equation}
    \text{E}_{\vec{\theta}_A'}[\vec{\nabla}C(\vec{\theta}_A')\cdot\hat{\vec{\ell}}]= \text{E}_{\vec{\theta}_A'}\left[\sum_l\partial_l C(\vec{\theta}_A') \hat{\ell}_l\right] = \sum_l\text{E}_{\vec{\theta}_A'}[\partial_l C(\vec{\theta}_A')] \hat{\ell}_l\,,
\end{equation}
and that $\text{E}_{\vec{\theta}_A'}\left[\sum_l\partial_l C(\vec{\theta}_A')\right]=0$ $\forall l$ (see Appendix C in~\cite{holmes2021connecting} for a proof of the last equality). 

We note that this result holds for both cases of the Lemma as we are only averaging over $\thv_A$. It therefore doesn't matter if $\thv_B$ is chosen randomly or as a deterministic offset from $\thv_A$. Additionally, the same argument holds if we instead average over $\thv_B$ as the average gradient over $\vec{\theta}_B'=\vec{\theta}_B-\ell\hat{\vec{\ell}}$ is also zero if $\vec{\theta}_B'$ is a random draw from the uniform distribution over the parameter space.

We note that if either $\thv_A$ or $\thv_B$ is chosen to be some fixed point, this proof fails. To see the problem, note that $\vec{\theta}_A'=\vec{\theta}_A+\ell\hat{\vec{\ell}}$ becomes a deterministic value in the limit $\ell\to 0$ or $\ell\to L$ if $\thv_A$ or $\thv_B$ is deterministic, respectively. The gradient therefore approaches a fixed value and the average gradient for a given $\ell$ is no longer guaranteed to be zero. In other words, we require $\vec{\theta}_A'$ to be a random draw from the uniform distribution over the parameter space for all values of $\ell$.

\section{Proof of Theorem~\ref{thm:grad->diffs}}\label{app:BP->NG}

We begin with the case that $\thv_B=\thv_A+L\hat{\vec{\ell}}$. The magnitude of the second moment of a difference $\Delta C=C(\vec{\theta}_B)-C(\vec{\theta}_A)$ in this case is
\begin{equation}
\begin{aligned} 
\big| \text{E}_{\vec{\theta}_A}[(\Delta C)^2]\big|
=&\Bigg|\int_{0}^{L}\int_{0}^{L}\text{E}_{\vec{\theta}_A}[(\vec{\nabla}C(\vec{\theta}_A+\ell\hat{\vec{\ell}})\cdot\hat{\vec{\ell}})(\vec{\nabla}C(\vec{\theta}_A+\ell'\hat{\vec{\ell}})\cdot\hat{\vec{\ell}})] d\ell d\ell'\Bigg|\\
=&\Bigg|\int_{0}^{L}\int_{0}^{L}\text{Cov}_{\vec{\theta}_A}\left((\vec{\nabla}C(\vec{\theta}_A+\ell\hat{\vec{\ell}})\cdot\hat{\vec{\ell}}),(\vec{\nabla}C(\vec{\theta}_A+\ell'\hat{\vec{\ell}})\cdot\hat{\vec{\ell}})\right) d\ell d\ell'\Bigg|\,.
\end{aligned}
\end{equation}
Note that we have dropped the square of the mean as the mean is zero, so the second moment is the covariance. Next, we upper bound this covariance using the product of the variances using the Cauchy-Schwartz inequality.  That is, given that $| \text{E}(A,B) | \leq \sqrt{\text{E}(A^2) \text{E}(B^2)}$, it follows that
\begin{equation}\label{eq:CovBound}
    |\text{Cov}(X,Y)| = |\text{E}(  (X - \mu_x) (Y - \mu_y)  ) | \leq \sqrt{\text{E}(  (X - \mu_x)^2 \text{E}(  (Y - \mu_y)^2}  = \sqrt{\text{Var}(X) \text{Var}(Y)} \, .
\end{equation}
Thus we can write,
\begin{equation}
\begin{aligned} 
\big| \text{E}_{\vec{\theta}_A}[(\Delta C)^2]\big|\le&\int_{0}^{L}\int_{0}^{L}\left(\text{Var}_{\vec{\theta}_A}(\vec{\nabla}C(\vec{\theta}_A+\ell\hat{\vec{\ell}})\cdot\hat{\vec{\ell}})\text{Var}_{\vec{\theta}_A}(\vec{\nabla}C(\vec{\theta}_A+\ell'\hat{\vec{\ell}})\cdot\hat{\vec{\ell}})\right)^\frac{1}{2} d\ell d\ell'\,.
\end{aligned}
\end{equation}
Then, we split these variances into the covariances between individual components of the gradient
\begin{equation}
\begin{aligned} 
\big| \text{E}_{\vec{\theta}_A}[(\Delta C)^2]\big|\le&\int_{0}^{L}\int_{0}^{L}\left(\sum_{i=1}^m\sum_{j=1}^m\sum_{p=1}^m\sum_{q=1}^m\text{Cov}_{\vec{\theta}_A}\left(\partial_i C(\vec{\theta}_A+\ell\hat{\vec{\ell}})\hat{\ell}_i,\partial_j C(\vec{\theta}_A+\ell\hat{\vec{\ell}})\hat{\ell}_j\right)\right. \\
&\,\,\,\,\,\,\left.\text{Cov}_{\vec{\theta}_A}(\partial_p C(\vec{\theta}_A+\ell\hat{\vec{\ell}})\hat{\ell}_p,\partial_q C(\vec{\theta}_A+\ell\hat{\vec{\ell}})\hat{\ell}_q)\right)^\frac{1}{2} d\ell d\ell'\,.
\end{aligned}
\end{equation}
As before, these covariances can be bounded by variances using the Cauchy-Schwartz inequality (Eq.~\eqref{eq:CovBound}),
\begin{equation}
\begin{aligned} 
\big| \text{E}_{\vec{\theta}_A}[(\Delta C)^2]\big|\le&\int_{0}^{L}\int_{0}^{L}\Bigg(\sum_{i=1}^m\sum_{j=1}^m\sum_{p=1}^m\sum_{q=1}^m\left(\text{Var}_{\vec{\theta}_A}\left(\partial_i C(\vec{\theta}_A+\ell\hat{\vec{\ell}})\right)\text{Var}_{\vec{\theta}_A}\left(\partial_j C(\vec{\theta}_A+\ell\hat{\vec{\ell}})\right)\right)^\frac{1}{2}\\
&\,\,\,\,\,\,\left(\text{Var}_{\vec{\theta}_A}\left(\partial_p C(\vec{\theta}_A+\ell\hat{\vec{\ell}})\right)
\text{Var}_{\vec{\theta}_A}\left(\partial_q C(\vec{\theta}_A+\ell\hat{\vec{\ell}})\right)\right)^\frac{1}{2}\Bigg)^\frac{1}{2} d\ell d\ell'\,.
\end{aligned}
\end{equation}
If each variance can be bounded by a function $F(n)$, we can then write
\begin{equation}\label{eq:mag_sq_bound}
\begin{aligned} 
\big| \text{E}_{\vec{\theta}_A}[(\Delta C)^2]\big|
\le&m^2\int_{0}^{L}\int_{0}^{L}F(n) d\ell d\ell'\\
=&m^2L^2F(n).
\end{aligned}
\end{equation}
We have thus established the first case in the theorem.

For the case where $\thv_B$ is also a random variable, the same analysis holds except that the $L$ in Eq.~\eqref{eq:mag_sq_bound} is a random variable. However, as by definition the cost functions we are working with are bounded, we can bound $L$ with a maximum distance $L\le L_\textrm{max}$. In that case we have
\begin{equation}
\begin{aligned} 
\big| \text{E}_{\vec{\theta}_A}[(\Delta C)^2]\big|
\le&m^2L^2F(n)\\
\le&m^2L_\textrm{max}^2F(n),
\end{aligned}
\end{equation}
which establishes the second case.

As before, we note that this result only holds when $\thv_A+\ell\hat{\vec{\ell}}$ represents a random draw from the uniform distribution at all points along the integration path. 

\section{Derivation of the Parameter Shift Rule}\label{app:E}

Here we recall the  derivation of the Parameter Shift Rule introduced in~\cite{mitarai2018quantum,schuld2019evaluating}. 
For simplicity, consider a simplified cost function of the form
\begin{equation}\label{eq:cost-SM}
    C(\thv)=\Tr[U(\thv)\rho U\ad(\thv)O]\,,
\end{equation}
where $\rho$ is an input quantum state, $U(\thv)$ a parametrized quantum circuit, and $O$ a Hermitian operator. The derivation for the general cost in the main text follows readily. We also recall that we have expressed $U(\thv)$ as
\begin{align}
    U(\thv)&=\prod_{j=1}^L e^{-i \theta_l H_l/2 }W_l
\end{align}
with $H_l$ an operator with only two non-zero eigenvalues normalized to $\pm 1$, and $W_l$ an unparametrized fixed unitary.

Then, one can trivially show that 
\begin{equation}
    \frac{\partial (e^{-i \theta_k H_k/2})}{\partial \theta_k}= -\frac{i}{2} H_k e^{-i \theta_k H_k/2}\,,
\end{equation}
so that 
\begin{equation}
    \frac{\partial U(\thv)}{\partial \theta_k}=\frac{-i}{2}U(\thv)_{L,j}H_k U(\thv)_{R,j}\,,
\end{equation}
with 
\begin{equation}
    U(\thv)_{L,j}=\prod_{l=j+1}^{m} e^{-i\theta_lH_l/2} W_l\,,\quad \text{and} \quad
    U(\thv)_{R,j}=\prod_{l=1}^{j} e^{-i\theta_lH_l/2} W_l\, .
\end{equation} 
Thus, a direct computation yields
\begin{align}
    \frac{\partial C(\thv)}{\partial \theta_k}&= \frac{\partial \Tr[U(\thv) \rho U\ad(\thv) O]}{\partial \theta_k}=\Tr[\frac{\partial U(\thv)}{\partial \theta_k}\rho U\ad(\thv) O] + \Tr[ U(\thv)\rho  \frac{\partial U\ad(\thv)}{\partial \theta_k}  O]\\
    &=-\frac{i}{2}\Tr\Big[U(\thv)_{L,j}[H_k, U(\thv)_{R,j}\rho U\ad(\thv)_{R,j}]U\ad(\thv)_{L,j} O\Big]\,.
\end{align}

Now, one can use the following useful inequality 
\begin{equation}\label{eq:lemma}
    i[H_k,A]=e^{i\pi H_k/4}A e^{-i\pi H_k/4}-e^{-i\pi H_k/4}A e^{i\pi H_k/4}\,.
\end{equation}
where $H_k$ is a Hermitian matrix with only two non-zero eigenvalues normalized to $\pm 1$ (i.e., $H_k$ is such that $H_k^2=\id$), and where $A$ is any matrix. Using Eq.~\eqref{eq:lemma} and noting that $e^{\pm i\pi H_k/4}e^{-i\theta_k H_k/2}=e^{-i(\theta_k\mp\pi/2) H_k/2}$ one finds 
\begin{equation}
    \partial_{\theta_j}C(\thv)=\frac{C(\thv'+\pi\hat{e}_j)-C(\thv')}{2}
\end{equation}
where $\hat{e}_j$ is the unit vector along the $j-th$ parameter direction and $\thv'=\thv-\pi/2\hat{e}_j$.

\section{Parameter dependence of variance in gradients}\label{app:paramdep}
\begin{figure}[ht]
    \centering
    \includegraphics[width=0.5\columnwidth]{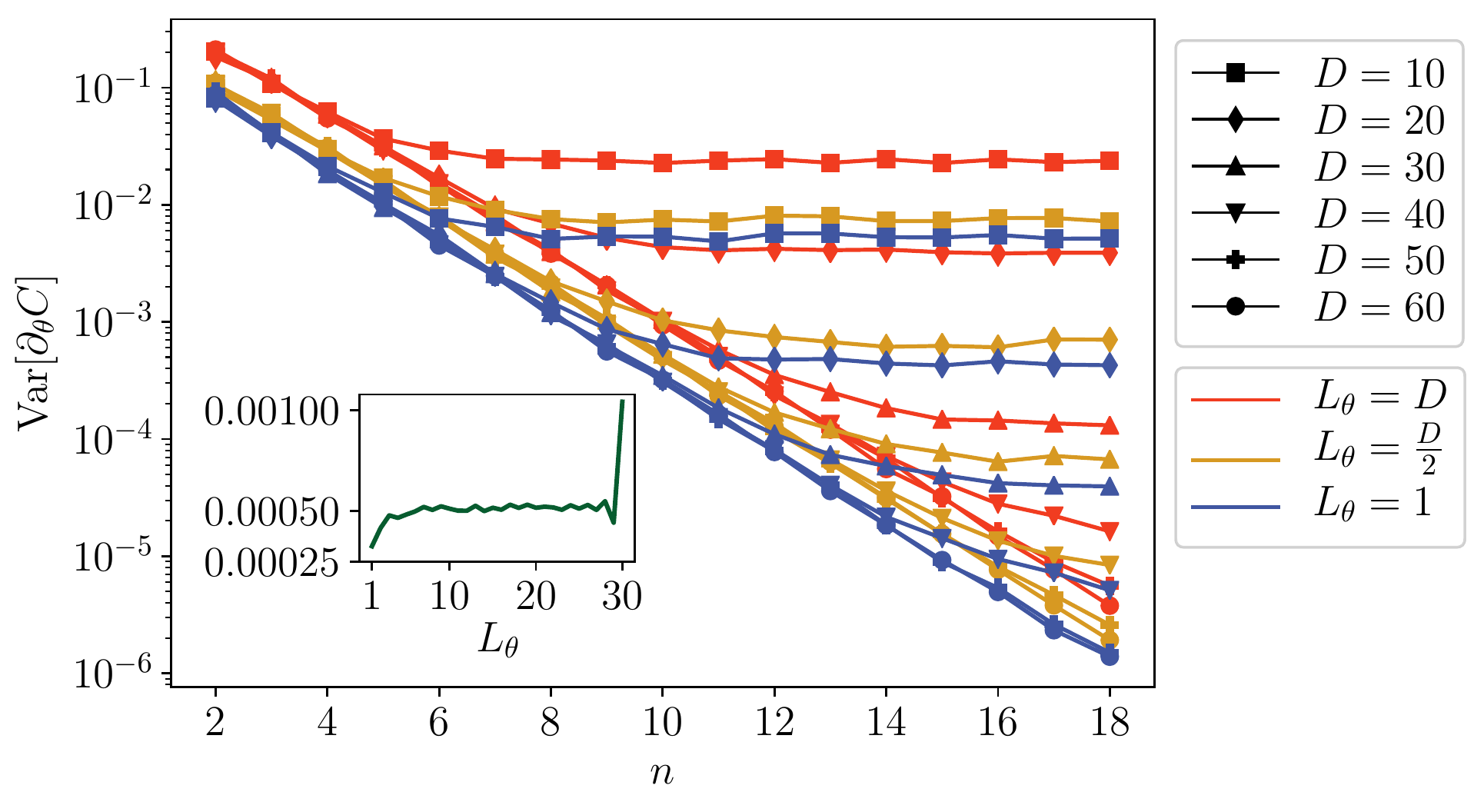}
    \caption{\textbf{Variance of partial derivative with respect to parameter at different depths in the PQC.} We show the scaling of the variance of partial derivatives the same system as in Fig.~\ref{fig:numerics}, but considering the partial derivatives of different parameters. We consider a parameter from each of the first, middle, and last layers of the PQC. For each choice of parameter location we find similar exponential scaling, though they do not always have the same prefactors. The inset plots the variance in the cost gradient as a function of $L_{
    \theta}$, the layer the derivative is taken with respect to. }
    \label{fig:layer_dependence}
\end{figure}

As mentioned in the main text, the variance of the partial derivatives of a cost function will in general depend on the parameter selected. We show an illustration of this dependence for the same landscape numerically explored in Section~\ref{sec:Numerics} in Fig.~\ref{fig:layer_dependence}. Since this cost function and PQC do exhibit a barren plateau we find that each case considered experiences exponential suppression, though with different prefactors.

\end{document}